\documentclass[twocolumn,showpacs,preprintnumbers,amsmath,amssymb,showkeys]{revtex4}

\usepackage{graphicx}
\usepackage{bm}

\begin{document}

\preprint{submitted to PRB}

\title{Decoupled CuO$_2$ and RuO$_2$ layers in superconducting and magnetically ordered 
RuSr$_2$GdCu$_2$O$_8$ }

\author{M.~Po\v{z}ek, A.~Dul\v{c}i\'{c}, D.~Paar, A.~Hamzi\'{c}, M.~Basleti\'{c}, E.~Tafra}
\affiliation{%
Department of Physics, Faculty of Science, University of Zagreb, P. O. Box 331,
HR-10002 Zagreb, Croatia
}%

\author{G.~V.~M.~Williams}
\affiliation{
2. Physikalisches Institut, Universit\"{a}t Stuttgart, D-70550 Stuttgart, Germany \\
and Industrial 
Research Limited, P.O. Box 31310, Lower Hutt, New Zealand.
}%
\author{S.~Kr\"{a}mer}
\affiliation{
2. Physikalisches Institut, Universit\"{a}t Stuttgart, D-70550 Stuttgart, Germany 
}%

\date{\today}

\begin{abstract}
Comprehensive measurements of dc and ac susceptibility, dc resistance, magnetoresistance, 
Hall resistivity, and microwave absorption and dispersion in fields up to 8~T have been 
carried out on RuSr$_2$GdCu$_2$O$_8$ with the aim to establish the properties of 
RuO$_2$ and CuO$_2$ planes. At $\sim$130~K, where the magnetic order develops in the 
RuO$_2$ planes, one observes a change in the slope of dc resistance, change in the 
sign of magnetoresistance, and the appearance of an extraordinary Hall effect. 
These features indicate that the RuO$_2$ planes are conducting. A detailed analysis 
of the ac susceptibility and microwave data on both, ceramic and powder samples show 
that the penetration depth remains frequency dependent and larger than the London 
penetration depth even at low temperatures. We conclude that the conductivity in 
the RuO$_2$ planes remains normal even when superconducting order is developed in 
the CuO$_2$ planes below $\sim$45~K. Thus, experimental evidence is provided in support 
of theoretical models which base the coexistence of superconductivity and magnetic 
order on decoupled CuO$_2$ and RuO$_2$ planes.
\end{abstract}

\maketitle

\section{\label{sec:level1}Introduction}

The coexistence of superconductivity and  magnetic order has placed the ruthenium 
cuprates in the focus of considerable work recently. \cite{Felner:97,Bernhard:99, 
McLaughlin:99,Tallon:00,Pickett:99,Chmaissem:00,Lynn:00,Fainstein:99,Williams:00, 
Pozek:01,Kumagai:01,Tokunaga:01} These superconductors were 
originally synthesized by Bauernfeind {\it et al.}. \cite{Bauernfeind:95,Bauernfeind:96}
Most recent reports have focused on RuSr$_2$RCu$_2$O$_8$ where R=Gd or Eu. 
Its crystal structure can be viewed as similar to that of the 
YBa$_2$Cu$_3$O$_7$ where the one dimensional (1D) CuO chains are replaced by 
two dimensional (2D) RuO$_2$ layers. Within this picture, it comes as no surprise that 
superconductivity 
may occur when the CuO$_2$ layers are properly doped, in analogy to other cuprate 
superconductors.  Recent X-ray absorption near edge structure (XANES)\cite{RSLiu:01}
and nuclear magnetic resonance (NMR) studies of RuSr$_2$RCu$_2$O$_8$ 
\cite{Kumagai:01,Tokunaga:01} revealed that ruthenium occurs in a mixed valence 
state as Ru$^{4+}$ and Ru$^{5+}$ with almost equal concentration. Thus, from the 
point of view of superconductivity, the role of RuO$_2$ planes is to act as the 
charge reservoir which is necessary to dope the superconducting CuO$_2$ planes.
 
One can interpret the crystal structure of RuSr$_2$RCu$_2$O$_8$ as CuO$_2$ layers which
are connected by perovskite ruthenate  SrRuO$_3$ via the apical oxygen atoms. \cite{Chen:01}
From this perspective, it comes as no surprise that magnetic ordering may occur in 
RuSr$_2$RCu$_2$O$_8$, as in most ruthenates of the Ruddlesden-Popper series 
Sr$_{n+1}$Ru$_n$O$_{3n+1}$.\cite{Ruddlesden:58} The most three dimensional member of 
the series is pseudocubic SrRuO$_3$ ($n$=$\infty$), which ferromagnetically orders at 
$T_m$=165~K.\cite{Randall:59,Kanbayashi:78}. The $n$=3 member Sr$_4$Ru$_3$O$_{10}$ is 
orthorhombic and becomes ferromagnetic below $T_m$=148~K.\cite{Cao:97} The effective 
dimensionality is drastically lowered in the $n$=2 member Sr$_3$Ru$_2$O$_7$. It shows 
magnetic correlations dominated by ferromagnetic instability  above $T^*$=17~K, and 
develops a canted antiferromagnetic instability below $T^*$. \cite{Ikeda:00,Liu:01} 
The 2D member Sr$_2$RuO$_4$ ($n$=1) does not order magnetically, and becomes superconducting 
at very low temperatures.\cite{Maeno:94}
 
The crystal structure of RuSr$_2$RCu$_2$O$_8$ has an additional complexity when compared 
to YBa$_2$Cu$_3$O$_7$. The RuO$_6$ octahedra in RuSr$_2$RCu$_2$O$_8$ are coherently rotated 
around the c-axis with domains extending up to 20~nm in diameter.\cite{McLaughlin:99} 
Rotations of the RuO$_6$ octahedra are common in the ruthenates and it is believed that 
the different magnetic order is due to structurally induced changes in the band structure. 
The rotation of the RuO$_6$ octahedra was observed in Sr$_3$Ru$_2$O$_7$,\cite{Huang:98} 
which shows competing, nearly degenerate magnetic instabilities.\cite{Ikeda:00,Perry:01} 
The importance of the rotation of the RuO$_6$ octahedra is best seen in 
Ca$_{2-x}$Sr$_x$RuO$_4$,\cite{Nakatsuji:00} which is the $n$=1 member of the 
Ruddlesden-Popper series with Ca substitution for Sr. 
Since Ca$^{2+}$ is smaller than Sr$^{2+}$, the substitution brings about a structural 
distortion in which the RuO$_6$ octahedra are rotated and flattened along the interlayer 
direction.\cite{Friedt:01} By varying the degree of the substitution, one obtains an 
intriguing phase diagram from paramagnetic metal to antiferromagnetic insulator. 
For some intermediate degrees of the substitution, one obtains a metallic system which 
shows an incomplete magnetic ordering at temperatures below $T_m$, and metamagnetic 
behavior similar to that observed in the $n$=2 member Sr$_3$Ru$_2$O$_7$. Hence, it is 
not surprising that the reported studies of the magnetic structure in RuSr$_2$RCu$_2$O$_8$ 
have shown some ambiguity. In some of the measurements of the zero field cooled (ZFC) 
dc susceptibility a clear ferromagnetic transition was observed,\cite{Kumagai:01} 
while in others a cusp-like signal characteristic of an antiferromagnetic transition 
was detected.\cite{Bernhard:99,Tallon:00,Williams:00} In all cases, though, 
there was a deviation of field cooled (FC) curves from ZFC ones, which proved the presence 
of a ferromagnetic component. Microscopic techniques could not resolve this ambiguity, 
either. For example, zero-field muon spin rotation study reported ferromagnetic order with the 
spontaneous magnetization in the ab plane.\cite{Bernhard:99} In contrast, neutron 
diffraction studies found evidence of antiferromagnetic order with the Ru moments 
aligned along the c axis.\cite{Lynn:00,Jorgensen:01,Takagiwa:01} The small ferromagnetic 
component was presumed to be produced by spin canting from the c axis. A recent 
magnetization study showed that the ferromagnetic component grows at higher 
fields.\cite{Williams:00} This provides evidence of a field induced transition which 
was attributed to a spin-flop transition. Similar field induced changes are observed 
in Sr$_3$Ru$_2$O$_7$ and some partially substituted Ca$_{2-x}$Sr$_x$RuO$_4$ samples with 
distorted RuO$_6$ octahedra. One should note, however, that the type of the magnetic 
order need not be simply related to the distortions of the RuO$_6$ octahedra. For example, 
it was found that the other interesting ruthenate cuprate 
RuSr$_2$R$_{2-x}$Ce$_x$Cu$_2$O$_{10+\delta}$ has the same distortion of the RuO$_6$ 
octahedra, as well as the same Ru-O-Ru and Ru-O-Cu bond lengths, found in 
RuSr$_2$RCu$_2$O$_8$.\cite{Williams:02} Yet, RuSr$_2$R$_{2-x}$Ce$_x$Cu$_2$O$_{10+\delta}$ 
is ferromagnetic while RuSr$_2$RCu$_2$O$_8$  is antiferromagnetically ordered at low fields. 
 
A number of studies on RuSr$_2$RCu$_2$O$_8$ have concluded that the RuO$_2$ layers are 
insulating and the transport properties are dominated by the CuO$_2$ layers. As mentioned above, 
it has also been concluded from a XANES study and NMR studies that Ru in the RuO$_2$ layers 
shows a mixed Ru valence, which has not been reported in other ruthenate compounds.  
This could also be understood within the model of insulating RuO$_2$ layers.  However, the 
magnetoresistance above the magnetic ordering temperature has a dependence on magnetic 
field that is not observed in the high temperature superconducting cuprates (HTSC),
\cite{McCrone:99} and clearly indicates that the transport process involves coupling 
to the Ru spins either from a conducting RuO$_2$ layer or via coupling between the 
CuO$_2$ layers and the spins in the RuO$_2$ layers.
 
In the present paper, we address the question of the coexistence of superconductivity 
and magnetic order in RuSr$_2$GdCu$_2$O$_8$. The question is reduced to the role played 
by CuO$_2$ and RuO$_2$ planes and their mutual couplings. We report our measurements of 
dc and ac susceptibility, dc resistance, magnetoresistance, Hall effect, and microwave 
absorption in RuSr$_2$GdCu$_2$O$_8$. The measurements have been done on the same sample 
prepared as sintered ceramic and powder diluted in epoxy resin. This facilitates the 
distinction between intergranular and intrinsic intragranular properties. We find evidence 
that RuSr$_2$GdCu$_2$O$_8$, as prepared in this study, has magnetic structure similar 
to Sr$_3$Ru$_2$O$_7$, and partially substituted Ca$_{2-x}$Sr$_x$RuO$_4$, which have 
no CuO$_2$ planes. Our results also show that the RuO$_2$ planes are 
conducting, but do not become superconducting. In other words, our observations are 
consistent with the picture in which the charge carriers in the CuO$_2$ and RuO$_2$ planes 
are decoupled.

\section{Experimental Details}

The RuSr$_2$GdCu$_2$O$_8$ ceramic samples were prepared from a stoichiometric mix of 
RuO$_2$, SrCO$_3$, Gd$_2$O$_3$ and CuO$_2$.  The powder was calcined in air at 960$^\circ$C 
for 10 hours and then pressed into pellets, which were sintered at 1010$^\circ$C for 10 hours 
to obtain the Sr$_2$GdCuO$_6$ and CuO$_2$ precursors. This process has been shown to 
prevent the formation of the SrRuO$_3$ impurity phase.  The compound was then sintered 
at 1050$^\circ$C in O$_2$ gas for 10 hours, 1055$^\circ$C in O$_2$ gas for 10 hours, 1060$^\circ$C 
in O$_2$ gas for 10 hours and finally 1060$^\circ$C in O$_2$ gas for 7 days.  The sample 
was reground after each sintering step.  The final processing has been shown to result 
in good quality samples where the transition into the bulk diamagnetic phase occurs 
for temperatures of up to 35~K.
 
It has been shown in our recent study on  RuSr$_2$EuCu$_2$O$_8$\cite{Pozek:01} that the electronic 
transport at low temperatures in the normal and superconducting states can be dominated by 
intergranular processes.  This has the effect of masking the intrinsic intragranular 
properties. For this reason, part of the sample was ground into a fine powder and then 
embedded in an epoxy resin.  Unfortunately, it has been found that it is not possible 
to align the ruthenate cuprates and hence we did not attempt to cure the resin in 
a magnetic field.
 
Resistivity, magnetoresistance and Hall effect measurements were done in the standard 
six-contact configuration using the rotational sample holder and the conventional ac 
technique (22~Hz, 1~mA), in magnetic fields up to 8 T. Temperature sweeps for the resistivity 
measurements were performed with carbon-glass and platinum thermometers, while magnetic 
field dependent sweeps were done at constant temperatures where the temperature was 
controlled with a capacitance thermometer.
 
The samples were characterized by both, dc and ac magnetization measurements 
using a SQUID magnetometer. The temperature dependent dc magnetization measurements 
were made in an applied magnetic field of 5~mT, while the ac susceptibility was measured in 
zero dc field with an ac field of 5~$\mu$T and a frequency of 1~kHz.
 
The microwave measurements were made in an elliptical $_e$TE$_{111}$ copper cavity 
operating at 9.3~GHz.  For the purpose of the present study it is essential to have a system 
 with high stability so that very small changes of the $Q$-factor can be reproducibly measured 
over long time scales. Therefore, the body of the microwave cavity was kept at liquid helium 
temperature. The unloaded cavity had a $Q$-factor of about 25\,000. The sample was mounted 
on a sapphire sample holder and positioned in the cavity center where the microwave electric 
field has maximum.  The temperature of the sample could be varied from liquid helium to room 
temperature. The cryostat with the microwave cavity was placed in a superconducting magnet 
so that the sample could be exposed to a dc magnetic field of up to 8 T.  The changes in the 
properties of the sample caused by either temperature variation or magnetic field were 
detected by a corresponding change in the $Q$-factor of the cavity and a resonant frequency shift. 
The quantity  $1/2Q$ represents the total losses of the cavity and the sample. The experimental 
uncertainty in the determination of  $1/2Q$ was about 0.03~ppm. We present our data as the 
difference $\Delta(1/2Q)$ between the measured values with and without the sample in the cavity.
In the case of the powder samples, the subtracted background signal was measured with a piece 
of clear epoxy of the same size as the sample with the powder. The resonant frequency of the 
cavity loaded with the sample was measured with a microwave frequency counter and the 
results are expressed as $\Delta f/f$ where $f$ is the frequency at the beginning 
of the measurement and $\Delta f$ is the frequency shift. The details of the detection 
scheme are given elsewhere \cite{Nebendahl:01}. 

In the present case, the microwave penetration depth is much less than the sample 
thickness and the measured quantities are simply related to the surface impedance of the 
material
\begin{equation}
Z_s=\sqrt{i \frac{\widetilde{\mu}_r \mu _0 \omega}{\widetilde{\sigma}}}
\label{eq:1}
\end{equation}
where  $\widetilde{\sigma}$ is the complex conductivity, and $\widetilde{\mu}_r$  is the complex relative permeability at the 
operating frequency. Both quantities can be temperature and field dependent.
The sample is placed in the center of the cavity where the magnetic component of  the 
microwave field has a node in the empty cavity. However, the wavelength in the conducting 
sample is much shorter than in vacuum so that a magnetic microwave field is also present 
within the skin depth from the sample surface.  The total microwave impedance comprises 
both nonresonant resistance and resonant spin contributions.

\section{Results and Analyses}

\subsection{Magnetization and ac susceptibility}

\begin{figure}
\includegraphics[width=20pc]{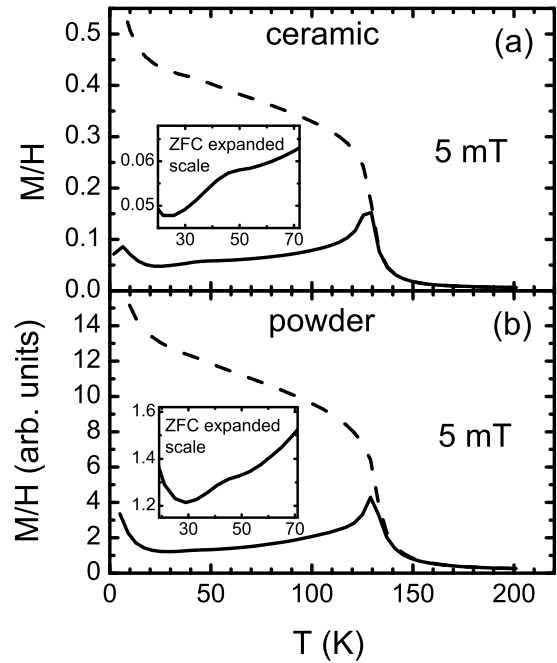}
\caption{Plots of the ZFC (solid curves) and FC (dashed curves) dc $M/H$ of polycrystalline 
(a) ceramic and (b) powder samples of RuSr$_2$GdCu$_2$O$_8$ in an applied field of 5~mT. The data 
has not been corrected for demagnetizing effects.} 
\label{fig_01}
\end{figure}

The dc magnetization curves observed in our RuSr$_2$GdCu$_2$O$_8$ samples are typical of 
those previously reported.\cite{Bernhard:99,Tallon:00,Williams:00}
Here we present in Fig. \ref{fig_01} the ZFC and FC dc 
magnetization at 5~mT in ceramic and powder samples taken from the same pellet. These 
curves show that the magnetic behavior observed in the ceramic sample is well reproduced in 
the powder sample, i.e. the influence of the intergranular medium on the dc-magnetic 
properties is negligible. There are three main features in ZFC curves: (i) a peak in the dc 
magnetization at $\sim$130~K, (ii) a decrease in the dc magnetization for temperatures less than 
$\sim$47~K, and (iii) an upturn of the magnetization below $\sim$20~K. The first feature is due to the 
magnetic ordering in the RuO$_2$ layers. The FC curves deviate strongly from the ZFC ones 
indicating that a ferromagnetic component is present in our samples, both ceramic and 
powder. The second feature near 47~K has been attributed to the superconducting transition, 
and the third feature below 20~K is due to the onset of the magnetic ordering of the Gd 
sublattice, which orders antiferromagnetically at 2.5~K.\cite{Bernhard:99,Lynn:00}

\begin{figure}
\includegraphics[width=20pc]{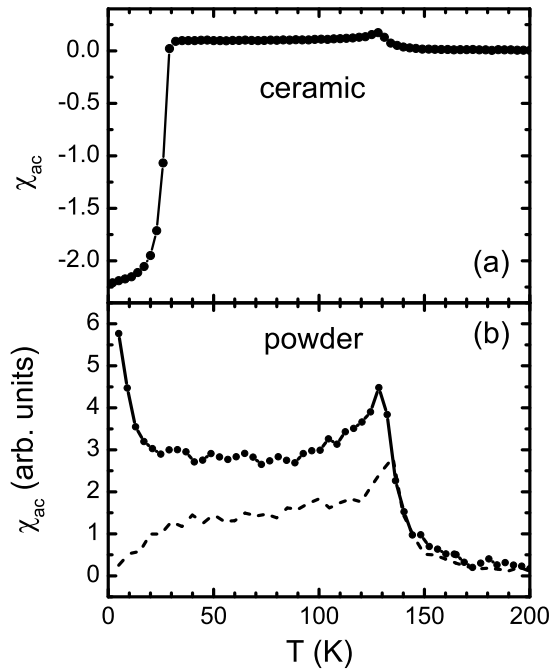}
\caption{Plots of ac susceptibilities of the same samples as in figure 1. The ac field 
amplitude was 5~$\mu$T, and frequency 1~kHz. The 
dashed line in (b) shows the ac susceptibility of a powder sample of  RuSr$_2$EuCu$_2$O$_8$. 
The data has not been corrected for demagnetizing effects.} 
\label{fig_02}
\end{figure}

The ac susceptibilities of the same ceramic and powder samples are shown in Fig. \ref{fig_02}. The magnetic ordering at $\sim$130~K is clearly seen in both, ceramic and powder samples. 
However, the superconducting transition, which is clearly seen in these samples below 47~K 
by dc magnetization in Fig. \ref{fig_01} is not manifested in the same way in the ceramic and powder 
samples when ac susceptibilities are measured. This is an unusual observation. The ac 
susceptibility curves in other HTSC exhibit nearly the same shapes and transition temperature 
widths for ceramic and powder samples of the same compound.
\cite{Athanassopoulou:96,Kirschner:99} In contrast, in Fig. \ref{fig_02}
we observe that a large diamagnetic shielding, which starts below 33~K in the ceramic 
sample, is not present in the powder. Hence, the large shielding signal in the ceramic sample 
could be interpreted as due to the onset of the intergranular Josephson currents. 
Superconductivity is certainly developed in the grains already below 47~K, but the 
intragranular ac screening currents appear to be very weak. They are so weak that even the 
high temperature tail of the Gd paramagnetic signal is sufficient to obscure their 
manifestation. In Fig. \ref{fig_02}b we present also the ac susceptibility of the powder sample of 
RuSr$_2$RCu$_2$O$_8$ with R=Eu, which is not paramagnetic. The intragranular superconducting 
signal is detectable in this compound. However, instead of showing a rapid drop just 
below the superconducting transition temperature
$T_c$, this signal exhibits a gradual decrease in the whole temperature range of the 
measurement. Obviously, the penetration depth in RuSr$_2$RCu$_2$O$_8$ does not drop 
rapidly from the normal state skin depth $\delta_n =\sqrt{2/\mu_0 \omega \sigma_n}$
to the London penetration depth  $\lambda_L$ as in other HTSC. For 
the operating frequency of 1~kHz, $\delta_n$  is typically much larger than the grain 
size ($\delta_n$ $\sim$1~cm). 
Below $T_c$, the ac conductivity becomes complex $\widetilde{\sigma}=\sigma_1 -i\sigma_2$, 
where the real and 
imaginary parts are due to the uncondensed normal electrons and the superconducting fluid, 
respectively. In the cuprate HTSC one reaches the condition $\sigma_1 \ll \sigma_2$ 
already a little below 
$T_c$. The penetration depth is then determined mainly by the superconducting fluid, 
and equals $\lambda_L$   independently of the operating frequency. 
Since  $\lambda_L$  is typically smaller than the grain 
size, the intragranular screening currents become effective. Therefore, the diamagnetic signal 
in the ac susceptibility is strong and follows the temperature dependence of $\lambda_L(T)$. 
In the 
case of powder  RuSr$_2$EuCu$_2$O$_8$we observe only a weak ac susceptibility signal. 
Below $T_c$ the penetration depth is reduced from its normal state value $\delta_n$ , 
but obviously not enough to 
become smaller than the grain size. We have to conclude that a large fraction of the charge 
carriers remains in the normal state at all temperatures below $\sim$45~K. The penetration 
depth at 1~kHz is then a combined effect of both, superconducting and normal electrons, and 
remains larger than the grain size. 
The magnetization results do not allow us to determine
the location of the normal state charge carriers at low temperatures below $T_c$. 
They are likely to be located either in the RuO$_2$ planes or in the CuO$_2$ planes. 
However, as we show later in this paper, it is possible to gather further information about
these low temperature normal state charge carriers and their location from microwave 
measurements.  

\subsection{DC resistance, magnetoresistance, and Hall  effect}

\begin{figure}
\includegraphics[width=20pc]{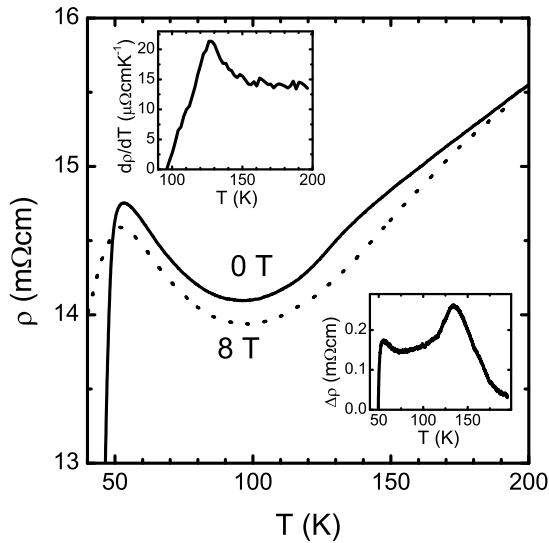}
\caption{Plots of dc resistivities of ceramic RuSr$_2$GdCu$_2$O$_8$ in
zero magnetic field (solid curve)
and in 8~T (dashed curve) field for transverse geometry 
($\mathbf{H}\perp\mathbf{I}$). The upper left inset shows the 
derivatives of the resistivity with respect to temperature, and 
the lower right inset shows the 
difference of the resistivities in zero magnetic field and in 8~T.}
\label{fig_03}
\end{figure}

The resistivity of RuSr$_2$GdCu$_2$O$_8$ has already been elaborated in some previous studies. 
\cite{Bernhard:99,Tallon:00} Here we focus 
on some features that have not been considered before and could elucidate the roles of CuO$_2$ 
and RuO$_2$ planes in the transport properties. Figure  \ref{fig_03} shows the resistivity curves in zero field 
and 8~T field. For the latter a transverse geometry was used ($\mathbf{H}\perp\mathbf{I}$). 
In general, the resistivity 
in ceramic samples may have contributions from intergranular medium and from intrinsic 
scattering process in the grains. We show below, using microwave measurements on a 
powder sample of RuSr$_2$GdCu$_2$O$_8$, that the resistivity in the grains does not exhibit a 
semiconducting contribution. Thus, the upturn of the resistivity below 100~K in 
Fig. \ref{fig_03} can 
be attributed to the prevalence of the intergranular contribution. Above 100~K, the 
predominant contribution to the total resistivity comes from the intragranular scattering. The 
relevant question is whether the whole charge transport occurs only in the CuO$_2$ planes, or 
there is an additional contribution of the RuO$_2$ planes to the total conductivity. The ZFC curve 
in Fig. \ref{fig_03} shows a more rapid decrease of the resistivity near the magnetic ordering 
temperature  $T_m$. This phenomenon is better dis\-played in the upper inset to 
Fig. \ref{fig_03} where the 
derivative of the resistivity with respect to temperature $d \rho/dT$ is seen to have a maximum 
at  $T_m$. A peak in  $d \rho/dT$ at  $T_m$ is commonly observed in $3d$ ferromagnetic conductors.
\cite{Campbell:82}
It was explained by Fisher and Langer \cite{Fisher:68} who considered the effect of short range 
fluctuations in the magnetization in ferromagnetic metals. A peak in $d\rho/dT$  was also 
observed in  SrRuO$_3$which is a $4d$ ferromagnet,\cite{Klein:96}  but the temperature dependence of   $d\rho/dT$
near  $T_m$ was different than that predicted by Fisher and Langer and observed in $3d$ 
ferromagnetic metals.
This deviation was ascribed to the bad metallicity of  
SrRuO$_3$. We may conclude that the observation of
 a peak in  $d \rho/dT$ in our ceramic RuSr$_2$GdCu$_2$O$_8$ is a 
clear sign that RuO$_2$ planes are conducting. At this point one can only list the factors which 
may influence the form of this peak. First, the magnetic order in RuSr$_2$GdCu$_2$O$_8$ is 
predominantly antiferromagnetic at low fields with only a small ferromagnetic component. 
Second, the magnetic scattering affects the charge carriers in the RuO$_2$ planes but need not 
have much influence on the conductivity in the CuO$_2$ planes. Finally, the total resistivity in 
Fig. \ref{fig_02} includes also the intergranular semiconducting contribution. It is not predominant at 
 $T_m$, but should not be totally neglected. For all these reasons, the form of the peak in the inset 
to Fig. \ref{fig_03} could deviate from that predicted by Fisher and Langer. 

The effect of an applied magnetic field of 8~T is to decrease the transverse resistivity 
for temperatures less than $\sim$200~K, which is opposite to the effect observed in the HTSC.
\cite{Harris:95} The general decrease in the resistivity at 8~T and for temperatures less than 200 
K can be attributed to a decrease in the spin scattering contribution to the resistivity within the 
RuO$_2$ layers due to ordering of the spins in the RuO$_2$ layers. The decrease in the resistivity at 
8~T is clearer in the lower insert to Fig. \ref{fig_03} where we plot the difference in the transverse 
resistivity against temperature.  The resulting peak reflects the effect of spin fluctuations 
which can be suppressed by the applied magnetic field. Similar behavior is observed in the 
ferromagnetic metals  SrRuO$_3$ and  Sr$_4$Ru$_3$O$_{10}$ as the magnetic field is increased.  
This provides additional evidence that the RuO$_2$ layers are conducting. 

\begin{figure}
\includegraphics[width=20pc]{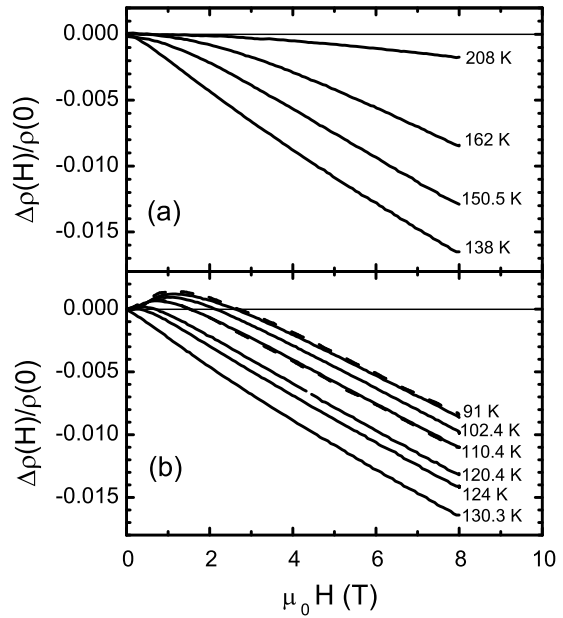}
\caption{Transverse magnetoresistance ($\mathbf{H}\perp\mathbf{I}$) 
$(\rho(H)-\rho(0))/\rho(0)$ in ceramic RuSr$_2$GdCu$_2$O$_8$ at 
various temperatures (a) above, and (b) below the magnetic ordering temperature. Also shown 
is the longitudinal magnetoresistance ($\mathbf{H}\parallel\mathbf{I}$) (dashed curves).} 
\label{fig_04}
\end{figure}
While the temperature dependence of the resistivity at 0~T and 8~T has features that are 
also observed in the ferromagnetic metals  SrRuO$_3$ and  Sr$_4$Ru$_3$O$_{10}$, 
we find that the transverse 
($\mathbf{H}\perp\mathbf{I}$) and longitudinal ($\mathbf{H}\parallel\mathbf{I}$) 
magnetoresistance in RuSr$_2$GdCu$_2$O$_8$ deviates from the behavior 
observed in  SrRuO$_3$ and  Sr$_4$Ru$_3$O$_{10}$.  This is 
apparent in Fig. \ref{fig_04} where we plot the transverse 
and longitudinal magnetoresistance above (Fig. \ref{fig_04}a) and below 
(Fig. \ref{fig_04}b) the magnetic ordering 
temperature.  Far above the magnetic ordering temperature, in the region where $M$ is 
proportional to $H$ (above 200~K), we find that $\Delta\rho_T /\rho_0 \propto M^\alpha$, 
where $\Delta\rho_T /\rho_0$  is the 
transverse magnetoresistance and $\alpha$=2. This observation is consistent with the results 
previously reported by McCrone {\it et al.}.\cite{McCrone:99}   
A similar magnetization dependence is observed 
in  SrRuO$_3$ and  Sr$_4$Ru$_3$O$_{10}$.  However, below 200~K we find that $\alpha$  
continously 
decreases to a value of $\alpha$=1 as the magnetic ordering temperature is approached.  
Furthermore, just below the magnetic 
ordering temperature, a positive transverse magnetoresistance is observed for low applied 
magnetic fields.  A low field positive transverse magnetoresistance is observed in 
SrRuO$_3$ but 
at a lower temperature and far below the ferromagnetic ordering temperature.\cite{Gausepohl:95}  In the case 
of  SrRuO$_3$, longitudinal magnetoresistance ($\mathbf{H}\parallel\mathbf{I}$) 
measurements do not reveal a positive magnetoresistance. 
\begin{figure}
\includegraphics[width=20pc]{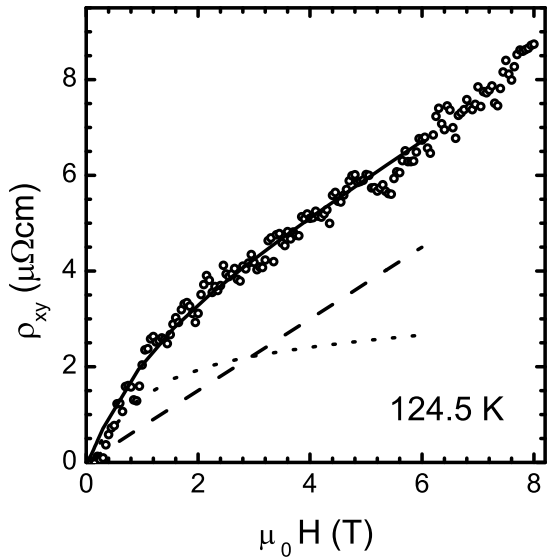}
\caption{Hall resistivity, $\rho_{xy}$, in ceramic RuSr$_2$GdCu$_2$O$_8$ at 124.5~K 
(circles).   Also 
shown is the ordinary Hall resistivity (dashed curve), the extraordinary Hall resitivity (dotted 
curve) and the total Hall resistivity as described by Eq.~(\ref{eq:2}) in the text.} 
\label{fig_05}
\end{figure}
Therefore, the positive transverse magnetoresistance observed in  SrRuO$_3$
was interpreted as being due to orbital magnetoresistance. The appearance of a low field 
positive magnetoresistance in both, transverse and longitudinal cases in Fig. \ref{fig_04}b 
calls for a 
different interpretation. It has been recently observed in Sr$_3$Ru$_2$O$_7$\cite{Liu:01} 
that below $T^*$=~17~K, 
where dc resistivity changes its slope and dc susceptibility exhibits a maximum, both, 
transverse and longitudinal magnetoresistance curves develop a positive low field 
contribution. The explanation was given in terms of magnetic instability present in this 
sample due to its distorted crystal structure. Similar features were found also 
in  Ca$_{2-x}$Sr$_x$RuO$_4$.\cite{Nakatsuji:00,Friedt:01}
It is possible that the positive transverse and longitudinal magnetoresistance observed 
in Fig. \ref{fig_04} are connected with the observation of antiferromagnetic 
and ferromagnetic order in 
RuSr$_2$GdCu$_2$O$_8$.  The simultaneous antiferromagnetic and ferromagnetic order in 
RuSr$_2$GdCu$_2$O$_8$ may be 
driven by band structure effects because the ferromagnetic component is different for 
different rare earths.  The low field ferromagnetic component is highest in RuSr$_2$YCu$_2$O$_8$ 
($\sim$30 \% of the AF component).

\begin{figure}
\includegraphics[width=20pc]{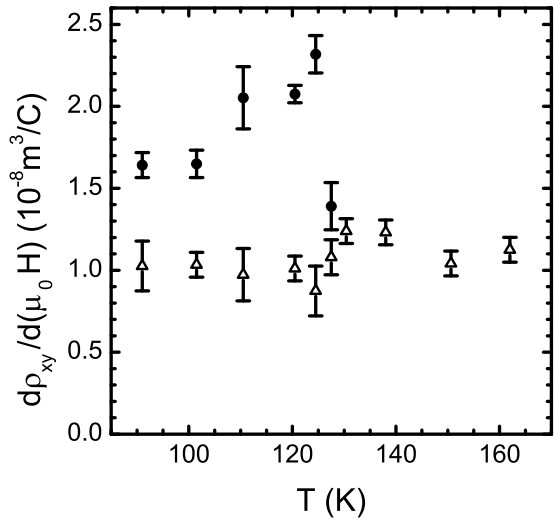}
\caption{Plot of the average $ d \rho_{xy}/d(\mu_0 H)$ against temperature in the low field region (filled 
circles) and in the high field region (open triangles) from ceramic RuSr$_2$GdCu$_2$O$_8$.} 
\label{fig_06}
\end{figure}
Besides having been useful in studying the nature of the magnetism in 
RuSr$_2$GdCu$_2$O$_8$, the observed magnetoresistance implicitly proves that RuO$_2$ planes are 
conducting, i.e. it supports further the conclusion reached from the analysis of the slope 
change in dc resistivity shown in Fig. \ref{fig_03}. Further evidence 
that the RuO$_2$ planes are 
conducting can be obtained from the analysis of the Hall effect in RuSr$_2$GdCu$_2$O$_8$.

Hall resistance, $\rho_{xy}$, was measured as a function of the applied magnetic field  $H$ at 
temperatures above and below  $T_m$. Fig. \ref{fig_05} shows 
the data at 124.5~K. For applied magnetic 
fields below 1~T a nonlinear increase of the Hall resistance $\rho_{xy}$ with increasing $H$ is observed. 
For high magnetic fields a linear field dependence of $\rho_{xy}$ dominates. The nonlinear 
increase in $\rho_{xy}$ with increasing $H$ is due to an additional term arising from the 
extraordinary Hall effect.  
This term is present in magnetic metals and it is due to skew scattering where the probability 
of scattering from $\mathbf k$ to $\mathbf {k'}$ is different from the probability of 
scattering from $\mathbf {k'}$ to $\mathbf {k}$.  The Hall 
effect in magnetic metals is commonly given by 
\begin{equation}
\rho_{xy}=R_0 \mu_0 H+R_s \mu_0 M
\label{eq:2},
\end{equation}
where $R_0$ is the 
ordinary Hall coefficient, $R_s$  is the extraordinary Hall coefficient,  $\mu_0$ is the vacuum 
permeability and $M$ is the magnetization of the sample. We show in Fig. \ref{fig_05} 
(solid curve) that Eq.~(\ref{eq:2}) 
does provide a reasonable representation of the data where $M$ 
from the 
RuO$_2$ layers was obtained from SQUID measurements.  Previous measurements of the Hall 
coefficient \cite{McCrone:99} were made at 8~T only, where it was found that the 
Hall coefficient displayed 
a peak near 160~K and decreased for temperatures less than 160~K.  However, the Hall 
coefficient measured in this way will be significantly affected by the anomalous Hall effect.  
The development of the anomalous Hall effect can be seen in Fig. \ref{fig_06} where we plot the 
average $d\rho_{xy} /d(\mu_0 H)$  for low and high fields.  It can be seen that the high field  
$d\rho_{xy} /d(\mu_0 H)$ is 
temperature independent and for $T<T_m$ the values are lower than those found 
from the average low field $d\rho_{xy} /d(\mu_0 H)$.  We note that the ordinary Hall effect 
contains contributions from both 
the CuO$_2$ and RuO$_2$ planes. We find that the high field  $d\rho_{xy} /d(\mu_0 H)$ 
is slightly greater than that 
observed in YBa$_2$Cu$_3$O$_{7-\delta}$ with a similar $T_c$ 
($R_H\sim 0.8\times 10^{-8}$ m$^3$C$^{-1}$)\cite{Segawa:01} while it is significantly greater than that observed in 
 SrRuO$_3$ ($R_H\sim 0.06\times 10^{-8}$ m$^3$C$^{-1}$)\cite{Klein:00}. This might suggest 
that the ordinary Hall effect in RuSr$_2$GdCu$_2$O$_8$ is dominated by the CuO$_2$ 
layers. However, the occurrence of the extraordinary Hall effect in RuSr$_2$GdCu$_2$O$_8$ 
indicates that the RuO$_2$ planes are conducting.

\subsection{Microwave measurements}

\begin{figure}
\includegraphics[width=20pc]{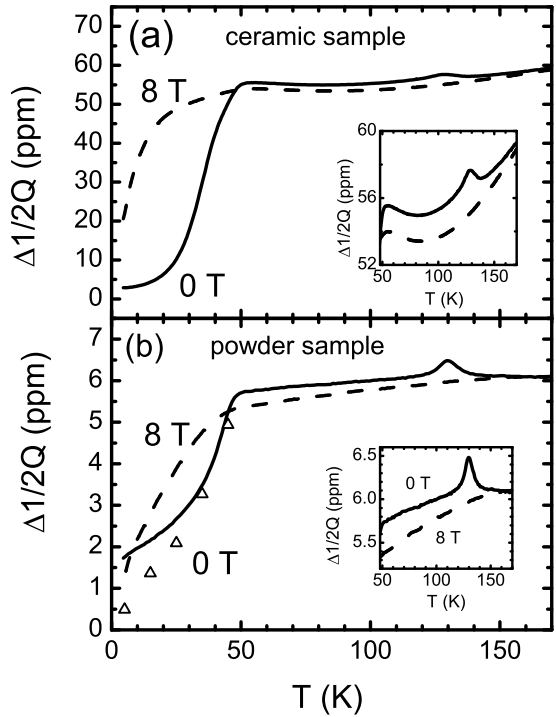}
\caption{Plot of $\Delta(1/2Q)$ in a (a) ceramic sample and a (b) powder sample of RuSr$_2$GdCu$_2$O$_8$
for applied magnetic fields of 0 (Earth field - solid curves) and 8~T (dashed curves). The 
inserts to (a) and (b) show an expanded view of $\Delta(1/2Q)$ at the two applied fields.  The open 
triangles in (b) show the zero field $\Delta(1/2Q)$ when the paramagnetic contribution of Gd$^{3+}$ ions 
is subtracted.} 
\label{fig_07}
\end{figure}

In Fig. \ref{fig_07} we plot the temperature dependences of $\Delta(1/2Q)$ for the ceramic and 
powder samples in zero and 8~T applied magnetic field. We first consider the microwave 
impedance in the normal state.  It is apparent in Fig. \ref{fig_07}a that the zero field 
$\Delta(1/2Q)$ of the 
ceramic sample shows a peak at the magnetic ordering transition temperature. Note that such 
a peak is not observed in the dc resistivity data of the same ceramic sample presented earlier, 
where a peak is observed only in the {\it{derivative}} $d\rho /dT$. When shown on an expanded scale in 
the insert to Fig. \ref{fig_07}a, this peak is seen to be superimposed on a decreasing resistive signal 
with some curvature due to the intergranular semiconducting medium. A similar peak was 
observed also in ceramic RuSr$_2$EuCu$_2$O$_8$ but the temperature dependence 
below the magnetic 
\begin{figure}
\includegraphics[width=20pc]{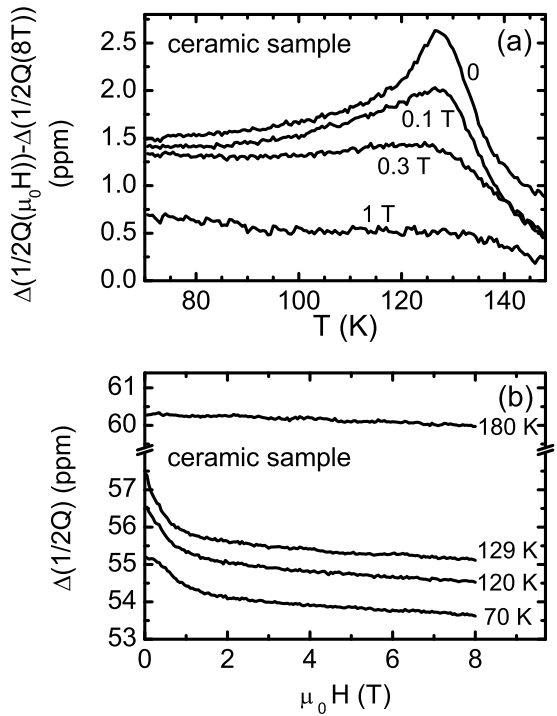}
\caption{(a) Temperature dependence of $\Delta(1/2Q)$ in the ceramic sample of RuSr$_2$GdCu$_2$O$_8$ at 
applied fields of 0, 0.1~T, 0.3~T, and 1~T after subtracting the signal taken at 8~T. (b) Magnetic 
field dependence of $\Delta(1/2Q)$ in the same sample at temperatures of 70~K, 120~K, 129~K and 
180~K. } 
\label{fig_08}
\end{figure}
ordering temperature was obscured by the microwave resistance from intergranular transport. 
The peak disappears at high magnetic fields as can be seen in Fig. \ref{fig_07}a where we plot 
$\Delta(1/2Q)$ at 8 T. We observe in Fig. \ref{fig_07}b that the semiconductor like upturn in 
$\Delta(1/2Q)$ of the 
ceramic sample is not present in the powder sample, thus providing clear evidence that it 
arises from intergranular conduction. However, the peak in $\Delta(1/2Q)$
is still seen in the powder 
sample. This peak is therefore an intrinsic property of the intragranular regions.  
It is possible that the peak arises from a change in $\widetilde{\mu}_r$ 
at the magnetic ordering temperature, although it can 
be seen in the insert to Fig. \ref{fig_07}b that this would require a large change in 
$\widetilde{\mu}_r$ over a small 
temperature range.  It may also be that model of Fisher and Langer for dc resistivity does not 
apply to microwave frequencies. We note that a peak in the dc resistance was predicted earlier 
by De Gennes and Friedel who based their calculation on the long range spin fluctuations.
\cite{DeGennes:58}

The suppression of the peak in $\Delta(1/2Q)$ by an applied magnetic field is evident in 
Fig. \ref{fig_08}a where we plot $\Delta(1/2Q)(\mu_0 H)-\Delta(1/2Q)(8\rm T)$.  
It is apparent that the peak rapidly 
disappears with increasing magnetic field and it vanishes completely at magnetic fields 
greater than 1 T.  Unlike the dc case, we find that the microwave magnetoresistance is 
negative for all temperatures in the normal state.  The magnetic field dependence of 
$\Delta(1/2Q)$ 
can be seen in Fig. \ref{fig_08}b.  For magnetic fields greater than $\sim$2~T, there is a linear decrease in 
$\Delta(1/2Q)$ with increasing magnetic field.  At 70~K and below, the Gd$^{3+}$ ESR absorption is 
evident in the low field region and centered near 0.3 T.  The intensity of this resonance 
increases with decreasing temperature owing to the increasing spin population difference in 
the lowest Gd$^{3+}$ spin levels.      
\begin{figure}
\includegraphics[width=20pc]{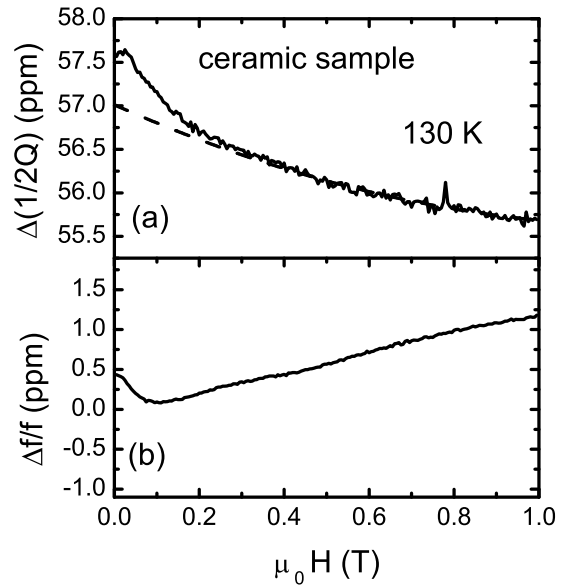}
\caption{Plot of $\Delta(1/2Q)$ (a) and $\Delta f/f$  (b) for magnetic fields of up to 1~T 
in the ceramic 
sample of RuSr$_2$GdCu$_2$O$_8$ at 130~K. The dashed curve shows the estimated absorption after 
subtraction of the contribution from the low field ferromagnetic resonance.} 
\label{fig_09}
\end{figure}

We show in Fig. \ref{fig_09} that there is an additional spin resonance below the magnetic 
ordering temperature. Here we plot $\Delta(1/2Q)$ and $\Delta f/f$ at 130~K 
and for magnetic fields of up to 
1 T.  For magnetic fields greater than $\sim$0.3 T, $\Delta(1/2Q)$ and $\Delta f/f$ have 
equal but opposite slopes 
as expected for a thick sample where microwave resistance is the only source of the 
microwave response.  However, at low fields one observes a peak in $\Delta(1/2Q)$ 
centered at $\sim$25 
mT, and $\Delta f/f$ displays a magnetic field dependence indicative of resonance phenomena.  
\begin{figure}
\includegraphics[width=20pc]{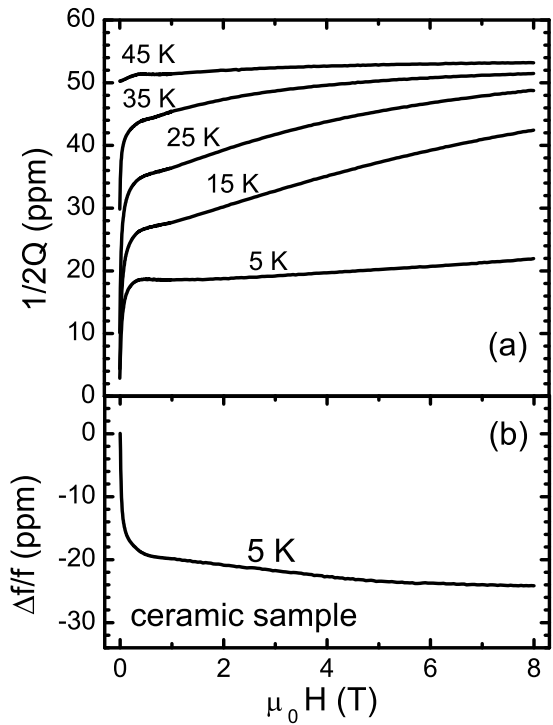}
\caption{(a) Field dependence of $\Delta(1/2Q)$ in the ceramic sample of RuSr$_2$GdCu$_2$O$_8$ at 
temperatures of 5~K, 15~K, 25~K, 35~K and 45~K.  The curves are taken after zero-field 
cooling. (b) Plot of $\Delta f/f$ at 5~K.} 
\label{fig_10}
\end{figure}
This feature could be due to the Ru ferromagnetic resonance observed by Fainstein et 
al.\cite{Fainstein:99} in RuSr$_2$GdCu$_2$O$_8$.  We estimate by the dashed curve 
in Fig. \ref{fig_09}a that this resonance contributes 
$\sim$0.7~ppm to $\Delta(1/2Q)$ at zero applied field.  However, it is apparent in 
Fig. \ref{fig_08} that this resonance contribution to $\Delta(1/2Q)$ is 
insufficient to explain 
the full height of the peak at  $T_m$.  
We note that $\Delta(1/2Q)$ and $\Delta f/f$ in the powder sample are similar to those 
in Figs. \ref{fig_08} and  \ref{fig_09} but the signal to noise ratio is much worse.

We now consider the microwave response in the superconducting state.  Returning to 
Fig. \ref{fig_07}, it can be seen that the zero field microwave resistance 
decreases near 50~K, similar 
to the dc case.  The effect of an applied magnetic field on  $\Delta(1/2Q)$ of 
the ceramic sample is 
similar to that observed in the HTSC.\cite{Pozek:92,Narlikar:96}  
However, the data for the powder sample is anomalous because at low temperatures the zero 
field values become larger than those taken 
at 8 T.  We show later that this behavior is due to an enhancement of  $\Delta(1/2Q)$ at 
low fields and low temperatures which is induced by the Gd$^{3+}$ resonance.  

\begin{figure}
\includegraphics[width=20pc]{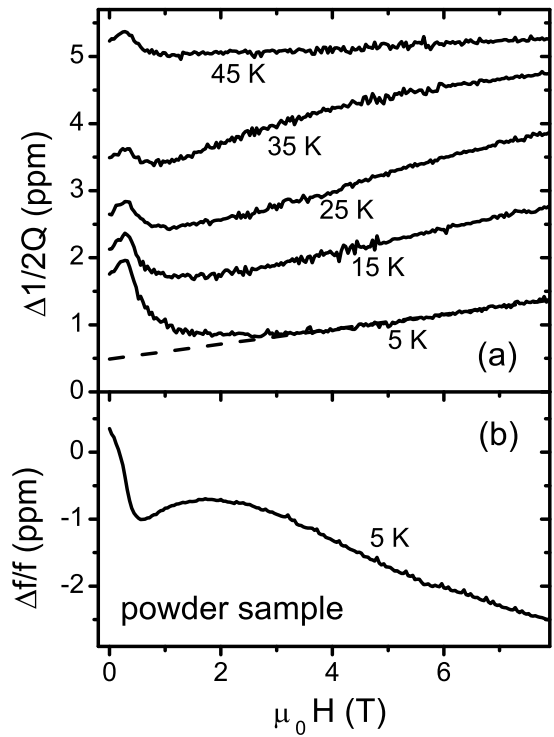}
\caption{(a) Field dependences of $\Delta(1/2Q)$ in the powder sample of RuSr$_2$GdCu$_2$O$_8$ at 
temperatures of 5~K, 15~K, 25~K, 35~K and 45~K.  The data was taken after zero-field cooling. 
The dashed line on the 5~K curve shows the extrapolation from higher fields to the zero field 
absorption value, which would remain after the subtraction of the Gd$^{3+}$ paramagnetic resonant 
absorption.  (b) Plot of $\Delta f/f$  for 5~K.} 
\label{fig_11}
\end{figure}
The magnetic field dependence of the microwave absorption in the superconducting 
state can be seen in Fig. \ref{fig_10}a where we plot  $\Delta(1/2Q)$ for temperatures of 5~K, 15~K, 25~K, 35 
K and 45~K.  The initial rapid increase in  $\Delta(1/2Q)$ is due to the Josephson coupled weak links 
that are being driven normal by the relatively small applied magnetic field.  The slower 
increase in  $\Delta(1/2Q)$ at higher magnetic fields arises from the absorption due to the increasing 
density of vortices in the grains.\cite{Coffey:91,Portis:93,Dulcic:93}
The effect of these processes on the frequency shift 
is seen in Fig. \ref{fig_10}b where  $\Delta f/f$ is plotted at 5~K. Above 35~K the Josephson coupling 
between grains becomes weaker. Thus, at 45~K no characteristic Josephson signal is seen in 
Fig. \ref{fig_10}a. At temperatures just below $T_c$, only the individual grains become superconducting. 
The intergranular coupling is established at a lower temperature.

Superimposed on the changes in  $\Delta(1/2Q)$ and  $\Delta f/f$ is the effect of the Gd$^{3+}$ resonance 
for low applied magnetic fields. It is partly obscured in Fig. \ref{fig_10} by 
the initial rapid changes 
in  $\Delta(1/2Q)$ and  $\Delta f/f$ induced by the weak link structure of the ceramic sample.  The effect of 
the Gd$^{3+}$ resonance is much clearer in Fig. \ref{fig_11} where we plot  
$\Delta(1/2Q)$ and  $\Delta f/f$ of the powder 
sample.  For applied magnetic fields above $\sim$2.5 T,  $\Delta(1/2Q)$ and  $\Delta f/f$ are dominated by the 
dissipative motion of the vortices in the mixed state \cite{Coffey:91,Portis:93,Dulcic:93}, while for magnetic fields less 
than $\sim$2.5~T the microwave response from Gd$^{3+}$ contributes significantly to  
$\Delta(1/2Q)$ and  $\Delta f/f$.   It is possible to account for the resonant contribution 
of Gd$^{3+}$ to  $\Delta(1/2Q)$ by extrapolating the 
high field  $\Delta(1/2Q)$ to the low field region as indicated by the dashed line in 
Fig. \ref{fig_11}.  By 
applying this correction, we show in Fig. \ref{fig_07}b (open triangles) that 
the microwave resistance 
at zero applied magnetic field is consistently smaller than that at 8~T. The absorption due to 
an effective microwave resistivity always increases with the applied magnetic field. There is, 
however, another unusual feature of the zero field microwave resistance in 
Fig. \ref{fig_07}b. Unlike 
the cuprate HTSC, where  $\Delta(1/2Q)$ drops rapidly below $T_c$ by more than two orders of 
magnitude, the zero field signal in Fig. \ref{fig_07}b is significant even for temperatures much less 
than $T_c$.

The anomalously large microwave resistance can not be accounted for by the 
occurrence of the spontaneous vortex phase. In this model, the ferromagnetic component of 
the spontaneous magnetization of the magnetically ordered RuO$_2$ layers generates 
vortices.\cite{Sonin:98}  The microwave currents would then induce oscillations of 
these vortices, leading to a 
microwave loss.\cite{Coffey:91,Portis:93,Dulcic:93}  The density of vortices, 
and hence the local field that is required for the 
increased  $\Delta(1/2Q)$ to be accounted for by the spontaneous vortex phase model, can be 
estimated from the observed rise of the signal level when the applied magnetic field is 
changed from zero to 8 T. We find that a spontaneous magnetic field of 7~T to 9~T is required.  
This is significantly larger than the local field estimated from a muon spin rotation study 
($\sim$0.1 T) or a Gd$^{3+}$ ESR study.\cite{Fainstein:99} The large microwave 
resistance at zero applied magnetic 
field is certainly due to a large fraction of the normal carriers still being present at 
temperatures well below $T_c$. Since the spontaneous vortex model is seen to be 
insufficient to provide the necessary amount of normal carriers, we propose that 
the normal carriers are to be found in the RuO$_2$ layers.  

At this point it is worthwhile to discuss whether the proposed 
interpretations of the ac susceptibility and microwave data are consistent. 
The operating frequencies for the 
two cases differ by seven orders of magnitude. The normal state skin depth is inversely 
proportional to the square root of the frequency. At our microwave frequency of 9.3~GHz, the 
skin depth in the normal state of RuSr$_2$GdCu$_2$O$_8$ is $\sim$5$\rm{\mu}$m, 
which is close to the grain size. As 
soon as a fraction of charge carriers is condensed into the superconducting state, the 
penetration depth at the microwave frequency is reduced below the grain size. This effect causes 
a significant drop of the signal level as shown in Fig. \ref{fig_07}b. This does 
not occur with the ac 
susceptibility signal in the same powder sample shown in Fig. \ref{fig_02}b. Due to 
the remaining 
fraction of the normal carriers, the penetration depth does not become frequency independent. 
The extremely weak superconducting signal inferred from Fig. \ref{fig_02}b implies that the 
penetration depth at 1~kHz is reduced from  $\delta_n$  $\sim$ 1 cm to a value still larger than the grain 
size. The comparison of the data at those two largely different frequencies provides the final 
proof that a large fraction of the charge carriers in RuSr$_2$GdCu$_2$O$_8$ is not condensed in the 
superconducting state even at very low temperatures. 

The proposed scenario would imply that there is no induced superconductivity in the 
RuO$_2$ layers.  This can be contrasted with fully loaded YBa$_2$Cu$_3$O$_7$, where the distance 
between the CuO$_2$ layer and the CuO chain is similar to the distance between the CuO$_2$ layer 
and RuO$_2$ layer in RuSr$_2$GdCu$_2$O$_8$, but there is induced superconductivity on the CuO chains 
in YBa$_2$Cu$_3$O$_7$. The idea of decoupled CuO$_2$ and RuO$_2$ planes has been mentioned in the early 
work of Felner et al.\ on a related ruthenate-cuprate compound 
RuSr$_2$Gd$_{1.4}$Ce$_{0.6}$Cu$_2$O$_{10-\delta}$.\cite{Felner:97} 
The conditions of decoupling have been treated theoretically.\cite{Pickett:99,Nakamura:01} 
The Ru $t_{2g}$ orbitals, where 
magnetism arises, are coupled to Cu $t_{2g}$ orbitals, but the latter are almost fully occupied. On 
the other hand, Ru $t_{2g}$ orbitals do not couple directly to the Cu $e_g$ orbitals, 
but only a more 
indirect coupling path via the apical oxygen may be possible. As a result, quite a small 
exchange splitting is induced in the antibonding $d_{x^2-y^2}-p_x$  ($dp\sigma$) 
orbitals in the CuO$_2$ 
planes. It was concluded that magnetism and superconductivity could coexist if the $dp\sigma$ 
orbitals formed the basis for superconductivity. The present paper provides experimental 
support for theoretical models based on decoupled subsystems in RuO$_2$ and CuO$_2$ planes.

\section{Conclusion}

        In conclusion, we find experimental evidence that, contrary to the conclusion from a 
number of previous studies, the RuO$_2$ layers in RuSr$_2$GdCu$_2$O$_8$ are conducting and contribute 
to the electronic transport above the superconducting transition. This is proven by the 
appearance of the peak in the temperature derivative of the dc resistivity $d\rho /dT$  at the 
magnetic ordering temperature  $T_m$, negative magnetoresistance, and extraordinary Hall 
resistivity. Hence, the insulating local moment model can not be applied to the RuO$_2$ layers in 
this compound. Rather, by combining the results of this study with NMR and XANES data 
which provide strong evidence of a mixed Ru valence,\cite{Kumagai:01,Tokunaga:01}
the RuO$_2$ planes can be 
described as conducting with a spatially varying charge density.        

The behavior below the 
superconducting transition is revealed from a detailed analysis of the dc magnetization, ac 
susceptibility, and microwave impedance data in the ceramic and powder samples. We prove 
that a large fraction of the charge carriers in RuSr$_2$GdCu$_2$O$_8$ is not condensed in the 
superconducting state even at temperatures far below $T_c$. The spontaneous vortex phase is 
found to be insufficient to account for the scale of the observed effect so that the normal 
conductivity is proposed to reside in the RuO$_2$ planes at all temperatures. The present paper 
provides experimental support for theoretical models which explain the coexistence of 
superconductivity and magnetism through effectively decoupled subsystems in CuO$_2$ and 
RuO$_2$ planes. 

\begin{acknowledgments}
We acknowledge funding support from the Croatian Ministry of Science and Technology, the 
New Zealand Marsden Fund and the Alexander von Humboldt Foundation.
\end{acknowledgments}

\end{document}